\documentclass[11pt]{article}

\IfFileExists{arxiv.sty}{\usepackage{arxiv}}{\usepackage[margin=1in]{geometry}}

\usepackage[utf8]{inputenc}
\usepackage[T1]{fontenc}
\usepackage{graphicx}
\usepackage{multirow}
\usepackage{amsmath,amssymb,amsfonts}
\usepackage{amsthm}
\usepackage{mathrsfs}
\usepackage[title]{appendix}
\usepackage{xcolor}
\usepackage{textcomp}
\usepackage{manyfoot}
\usepackage{booktabs}
\usepackage{algorithm}
\usepackage{algorithmicx}
\usepackage{algpseudocode}
\usepackage{listings}
\usepackage{bm}
\usepackage{microtype}
\usepackage{url}
\usepackage{hyperref}
\usepackage{natbib}

\theoremstyle{plain}

\theoremstyle{remark}

\theoremstyle{definition}

\providecommand{\shorttitle}{}
\providecommand{\keywords}[1]{\par\noindent\textbf{Keywords:} #1\par}
\title{Robust Deep Mixture Models}
\author{
Jinran Wu\\
School of Mathematics and Physics\\
The University of Queensland\\
St Lucia, Queensland 4072, Australia\\
\texttt{jinran.wu@uq.edu.au}
\and
\textbf{Geoffrey J. McLachlan}\thanks{Corresponding author: \texttt{g.mclachlan@uq.edu.au}}\\
School of Mathematics and Physics\\
The University of Queensland\\
St Lucia, Queensland 4072, Australia\\
\texttt{g.mclachlan@uq.edu.au}
}

\renewcommand{\shorttitle}{Robust Deep Mixture Models}

\begin{document}
\maketitle

\begin{abstract}
We propose a robust deep mixture model based on a pathway-wise shared scale-mixture construction. Layer-specific component indicators are independently distributed according to their corresponding mixing proportions and jointly define a complete pathway through the latent hierarchy. Conditional on the selected pathway, a single gamma-distributed latent precision variable is shared across the deepest latent distribution, every intermediate latent transition, and the observation model. Integrating out this shared precision yields an exact multivariate Student-$t$ distribution for each complete pathway, allowing robustness to propagate coherently throughout the entire latent hierarchy rather than being introduced separately within individual latent layers. Model parameters are estimated using a stochastic expectation--maximisation algorithm. Complete-pathway responsibilities are evaluated analytically, whereas the shared latent precision variables and latent Gaussian variables are generated from their conditional distributions before updating the model parameters. The pathway-specific degrees-of-freedom parameters are estimated by one-dimensional numerical optimisation. Simulation studies demonstrate accurate recovery of the pathway-specific degrees-of-freedom parameters together with consistently improved clustering performance relative to the deep Gaussian mixture model under heavy-tailed and contaminated settings. Real-data applications further illustrate the ability of the proposed model to identify heterogeneous latent structures while reducing the influence of atypical observations. The proposed framework retains the hierarchical representation and parsimonious parameter-sharing structure of the deep Gaussian mixture model while providing coherent pathway-wise robustness.
\end{abstract}


\section{Introduction}

Recent advances in deep learning have demonstrated the effectiveness of hierarchical representation learning for modelling complex, high-dimensional data \citep{lecun2015deep, goodfellow2016deep}. Motivated by these developments, probabilistic deep latent-variable models have attracted increasing attention because they combine hierarchical representations with statistically principled generative modelling. Unlike deterministic deep neural networks, probabilistic latent-variable models provide interpretable latent representations together with principled uncertainty quantification, making them particularly attractive for unsupervised learning tasks such as clustering, density estimation, and representation learning.

Finite mixture models are among the most widely used probabilistic frameworks for model-based clustering and density estimation \citep{mclachlan2000finite, mclachlan2019finite}. By representing a heterogeneous population as a mixture of relatively simple component distributions, they provide flexible density models with interpretable latent class structures. Classical Gaussian mixture models, however, become difficult to estimate in high-dimensional settings because unrestricted component covariance matrices require a rapidly increasing number of parameters. Mixtures of factor analysers (MFA) address this limitation by embedding a factor-analytic representation within each mixture component and replacing unrestricted covariance matrices with low-dimensional latent structures \citep{fokoue2003mixtures, mclachlan2003modelling}. This parameterisation substantially reduces the number of covariance parameters while allowing each component to retain its own local dependence structure. Consequently, MFA provides a flexible and parsimonious framework for clustering and density estimation of high-dimensional continuous data.

A parallel line of research has focused on robustness. Gaussian mixture models and MFA can be highly sensitive to outliers and heavy-tailed observations because observations with large Mahalanobis distances may exert excessive influence on maximum-likelihood estimation. Mixture models based on the multivariate Student-$t$ distribution address this problem through the Gaussian scale-mixture representation of the $t$ distribution, where observation-specific latent precision variables automatically down-weight atypical observations during estimation \citep{mclachlan1998robust, peel2000robust}. This approach has subsequently been extended to mixtures of $t$ factor analysers (MtFA), combining robustness and dimension reduction within a unified latent-variable framework \citep{mclachlan2007extension, baek2011mixtures, wang2022robust}. Alternative heavy-tailed formulations, including mixtures based on truncated $t$ distributions, have further expanded the class of robust mixture models \citep{wang2026robust, sheng2026robust}.

Although MFA provides an effective shallow latent representation, many high-dimensional datasets exhibit hierarchical dependence structures that cannot be adequately represented by a single latent layer. Motivated by hierarchical representation learning, \citet{viroli2019deep} introduced the deep Gaussian mixture model (DGMM), which recursively stacks factor-analytic mixture layers to construct a hierarchy of latent variables. This architecture increases modelling flexibility while retaining the parsimonious covariance structure inherited from factor analysis. The DGMM framework has subsequently been extended to variational Bayesian inference for high-dimensional data \citep{kock2022variational}, mixed-data modelling \citep{fuchs2022mixed}, missing-data imputation \citep{fuchs2022mi2ami}, topic modelling \citep{viroli2021deep}, and small-sample supervised learning \citep{gorshenin2025small}. Related deep probabilistic mixture architectures have also been investigated \citep{hamalainen2020deep}, while empirical studies have highlighted the multimodality of the DGMM likelihood and its sensitivity to initialisation \citep{selosse2020bumpy}. Recent applications further demonstrate the potential of deep mixture models in image analysis and other high-dimensional learning problems \citep{rakotonirina2026unsupervised,mahdavi2026image}. Despite these developments, existing DGMM formulations remain fundamentally Gaussian and therefore inherit the sensitivity of Gaussian latent-variable models to atypical observations.

Robust latent-variable mixture models and deep probabilistic mixture models have therefore evolved largely along separate research directions. Robust mixture models achieve resistance to atypical observations through heavy-tailed component distributions, but their latent representations are typically limited to shallow architectures. Conversely, existing deep mixture models provide flexible hierarchical representations but almost exclusively rely on Gaussian assumptions. More importantly, existing robust formulations introduce robustness locally within individual latent layers rather than propagating it coherently throughout the entire latent hierarchy. To the best of our knowledge, no existing probabilistic framework combines the hierarchical representation of the DGMM with pathway-wise robustness induced by an exact multivariate Student-$t$ construction.

To address these limitations, we propose a robust deep mixture model (RDMM) based on a pathway-wise shared scale-mixture construction. As in the original DGMM, component indicators at different latent layers are independently distributed according to layer-specific mixing proportions, and together define a complete pathway through the latent hierarchy. Conditional on the selected pathway, a single gamma-distributed latent precision variable is introduced for each observation and shared across the deepest latent distribution, every intermediate latent transition, and the observation model. Integrating out this shared precision yields an exact multivariate Student-$t$ distribution for each complete pathway, rather than introducing heavy-tailed behaviour separately within individual latent layers. Consequently, observations that are poorly explained by a given pathway receive small posterior precision weights and are automatically down-weighted throughout the entire latent hierarchy. The proposed model therefore preserves the hierarchical representation, parsimonious parameter-sharing structure, and dimension-reduction capability of the DGMM while providing coherent robustness against heavy-tailed observations and contamination.

The contributions of this paper are threefold. First, we introduce a novel pathway-wise shared scale-mixture formulation for deep mixture models. Unlike existing robust latent-variable models that employ independent scale variables within individual latent layers, the proposed model shares a single observation-level latent precision variable across the entire hierarchical pathway. This construction yields exact pathway-wise multivariate Student-$t$ distributions while providing coherent robustness throughout the latent hierarchy. Second, we develop a stochastic expectation--maximisation algorithm tailored to the proposed hierarchical scale-mixture model. Complete-pathway responsibilities are evaluated analytically, whereas the shared latent precision variables and latent Gaussian variables are generated from their conditional distributions. Model parameters, including the pathway-specific degrees-of-freedom parameters, are subsequently updated using stochastic sufficient statistics. Third, the proposed framework retains the hierarchical representation and parsimonious parameter-sharing structure of the DGMM while substantially improving robustness under heavy-tailed and contaminated data. Simulation studies demonstrate accurate recovery of the pathway-specific degrees-of-freedom parameters together with consistently improved clustering performance relative to the DGMM, while real-data applications further illustrate the practical effectiveness of the proposed approach.

\section{Robust deep mixture model formulation}
\label{sec:rdmm-t-model}

We now introduce the proposed RDMM, which extends the DGMM by incorporating pathway-specific multivariate Student-$t$ distributions. The proposed model combines the hierarchical latent representation of the DGMM with the Gaussian scale-mixture representation of the multivariate $t$ distribution. It therefore provides robustness against outlying or contaminated observations while preserving the dimension-reduction properties and local parameter-sharing structure of deep latent-variable models.

Consider observations $\bm{y}_j\in\mathbb{R}^{p}$, $j=1,\ldots,n$, generated from a hierarchical latent-variable model with progressively decreasing latent dimensions
\[
p=r_0>r_1>\cdots>r_h\geq 1,
\]
where $h$ denotes the number of latent layers and $\bm{z}_j^{(l)}\in\mathbb{R}^{r_l}$ denotes the latent vector associated with observation $j$ at layer $l$. For notational convenience, define $\bm{z}_j^{(0)}=\bm{y}_j. $

Conditional on a complete pathway through the hierarchy, the proposed model introduces a single positive latent precision variable shared across all layers. Consequently, the deepest latent distribution, every latent transition, and the observation equation are jointly scaled by the same random precision. This pathway-wise construction propagates robustness coherently throughout the hierarchy and yields an exact multivariate Student-$t$ marginal distribution for each complete pathway after the shared precision variable is integrated out.

Throughout this paper, the gamma distribution is parameterised in terms of shape and rate. Specifically,
\[
W\sim\operatorname{gamma}(a,b)
\]
has density
\[
f(w)
=
\frac{b^a}{\Gamma(a)}w^{a-1}\exp(-bw),
\qquad w>0,
\]
and mean
$\mathbb{E}(W)=a/b.$

\subsection{Hierarchical pathway structure}
\label{sec:rdmm-t-pathway}

Let $K_l$ denote the number of mixture components at layer $l$, and let
\[
S_j^{(l)}\in\{1,\ldots,K_l\}
\]
denote the component indicator for observation $j$ at that layer. The layer-specific component indicators are assumed to be independent a priori, with
\begin{equation}
\Pr\!\left(S_j^{(l)}=a\right)
=
\pi_a^{(l)},
\qquad
a=1,\ldots,K_l,
\qquad
\sum_{a=1}^{K_l}\pi_a^{(l)}=1,
\label{eq:rdmm-t-layerprob}
\end{equation}
for $l=1,\ldots,h$.

Define the random pathway indicator by
\[
\bm{S}_j
=
\left(S_j^{(1)},\ldots,S_j^{(h)}\right),
\]
and let
\[
\bm{s}=(s_1,\ldots,s_h)\in\mathcal{S},
\qquad
\mathcal{S}
=
\prod_{l=1}^{h}\{1,\ldots,K_l\},
\]
denote a generic realisation of a complete pathway. By the prior independence of the layer-specific component indicators, the prior probability of pathway $\bm{s}$ is
\begin{equation}
\pi_{\bm{s}}
=
\Pr(\bm{S}_j=\bm{s})
=
\prod_{l=1}^{h}\pi_{s_l}^{(l)},
\qquad
\sum_{\bm{s}\in\mathcal{S}}\pi_{\bm{s}}=1.
\label{eq:rdmm-t-pathprob}
\end{equation}

This factorisation retains the parsimonious mixing-proportion structure of the DGMM. In particular, the number of free mixing-proportion parameters is
\[
\sum_{l=1}^{h}(K_l-1),
\]
rather than one unrestricted parameter for each of the $\prod_{l=1}^{h}K_l$ complete pathways. Although the layer indicators are independent a priori, they are generally dependent conditional on $\bm{y}_j$, because the likelihood associated with a complete pathway depends jointly on all layer-specific component indices.

\subsection{Local layer-component parameterisation}
\label{sec:rdmm-t-local-parameterization}

In the baseline formulation, the parameters at layer $l$ are indexed only by the local component $s_l$. Specifically, they are
\[
\bm{\eta}^{(l)}_{s_l},
\qquad
\bm{\Lambda}^{(l)}_{s_l},
\qquad
\bm{\Psi}^{(l)}_{s_l}.
\]
Consequently, pathways containing the same local component at layer $l$ share the corresponding location vector, loading matrix, and specific covariance matrix. 

A more general parent-specific parameterisation could instead index these parameters by the pathway prefix
\[
\bm{s}_{1:l}=(s_1,\ldots,s_l),
\]
so that
\[
\bm{\eta}^{(l)}_{\bm{s}_{1:l}},
\qquad
\bm{\Lambda}^{(l)}_{\bm{s}_{1:l}},
\qquad
\bm{\Psi}^{(l)}_{\bm{s}_{1:l}}.
\]
This extension is not considered in the baseline model because it would substantially increase the number of parameters and weaken the sharing of information across pathways.

\subsection{Shared scale-mixture representation and generative process}
\label{sec:rdmm-t-generative}

Conditional on the selected pathway $\bm{S}_j=\bm{s}$, introduce the shared latent precision variable
\begin{equation}
W_{j\bm{s}}
\mid
\bm{S}_j=\bm{s}
\sim
\operatorname{gamma}\!\left(
\frac{\nu_{\bm{s}}}{2},
\frac{\nu_{\bm{s}}}{2}
\right),
\qquad
\mathbb{E}\!\left(W_{j\bm{s}}\mid\bm{S}_j=\bm{s}\right)=1,
\qquad
\nu_{\bm{s}}>2.
\label{eq:rdmm-t-W}
\end{equation}
The restriction $\nu_{\bm{s}}>2$ ensures that the pathway-level covariance matrix is finite.

The same scalar $W_{j\bm{s}}$ is shared across every level of the selected pathway. Thus, for a given observation and pathway, all latent innovations and the observation-level error are simultaneously inflated or deflated. A small posterior value of $W_{j\bm{s}}$ indicates that the observation is poorly explained by the selected pathway and reduces its influence on parameter estimation at every layer.

Conditional on the selected pathway and the shared precision variable, the deepest latent vector is generated according to
\begin{equation}
\bm{z}_j^{(h)}
\mid
\bm{S}_j=\bm{s},
W_{j\bm{s}}
\sim
\mathcal{N}_{r_h}\!\left(
\bm{0},
\frac{\bm{I}_{r_h}}{W_{j\bm{s}}}
\right).
\label{eq:rdmm-t-deepest}
\end{equation}

For $l=h,h-1,\ldots,2$, the latent transitions are
\begin{equation}
\bm{z}_j^{(l-1)}
\mid
\bm{z}_j^{(l)},
\bm{S}_j=\bm{s},
W_{j\bm{s}}
\sim
\mathcal{N}_{r_{l-1}}\!\left(
\bm{\eta}^{(l)}_{s_l}
+
\bm{\Lambda}^{(l)}_{s_l}\bm{z}_j^{(l)},
\frac{\bm{\Psi}^{(l)}_{s_l}}{W_{j\bm{s}}}
\right).
\label{eq:rdmm-t-latent}
\end{equation}
The observation equation is
\begin{equation}
\bm{y}_j
\mid
\bm{z}_j^{(1)},
\bm{S}_j=\bm{s},
W_{j\bm{s}}
\sim
\mathcal{N}_{p}\!\left(
\bm{\eta}^{(1)}_{s_1}
+
\bm{\Lambda}^{(1)}_{s_1}\bm{z}_j^{(1)},
\frac{\bm{\Psi}^{(1)}_{s_1}}{W_{j\bm{s}}}
\right).
\label{eq:rdmm-t-y}
\end{equation}

For each layer $l=1,\ldots,h$, the parameters have dimensions
\[
\bm{\eta}^{(l)}_{s_l}\in\mathbb{R}^{r_{l-1}},
\qquad
\bm{\Lambda}^{(l)}_{s_l}\in\mathbb{R}^{r_{l-1}\times r_l},
\qquad
\bm{\Psi}^{(l)}_{s_l}\in\mathbb{R}^{r_{l-1}\times r_{l-1}},
\]
where $r_0=p$. Each $\bm{\Psi}^{(l)}_{s_l}$ is positive definite and is commonly restricted to be diagonal for parsimony.

\subsection{Pathway marginals and observed-data likelihood}
\label{sec:rdmm-t-likelihood}

For a fixed pathway $\bm{s}=(s_1,\ldots,s_h)$, define its mean vector and scale matrix recursively from the deepest layer to the observation layer. Set
\[
\bm{\mu}_{\bm{s}}^{(h)}=\bm{0},
\qquad
\bm{\Sigma}_{\bm{s}}^{(h)}=\bm{I}_{r_h}.
\]
For $l=h,h-1,\ldots,1$, define
\begin{align}
\bm{\mu}_{\bm{s}}^{(l-1)}
&=
\bm{\eta}_{s_l}^{(l)}
+
\bm{\Lambda}_{s_l}^{(l)}\bm{\mu}_{\bm{s}}^{(l)},
\label{eq:rdmm-t-mu-recursion}
\\
\bm{\Sigma}_{\bm{s}}^{(l-1)}
&=
\bm{\Psi}_{s_l}^{(l)}
+
\bm{\Lambda}_{s_l}^{(l)}
\bm{\Sigma}_{\bm{s}}^{(l)}
\bm{\Lambda}_{s_l}^{(l)\top}.
\label{eq:rdmm-t-sigma-recursion}
\end{align}
The resulting observation-level pathway mean and scale matrix are
\[
\bm{\mu}_{\bm{s}}=\bm{\mu}_{\bm{s}}^{(0)},
\qquad
\bm{\Sigma}_{0,\bm{s}}=\bm{\Sigma}_{\bm{s}}^{(0)}.
\]

Because the same precision variable scales every level of the hierarchy, the conditional pathway marginal for the observation vector is
\begin{equation}
\bm{y}_j
\mid
\bm{S}_j=\bm{s},
W_{j\bm{s}}
\sim
\mathcal{N}_{p}\!\left(
\bm{\mu}_{\bm{s}},
\frac{\bm{\Sigma}_{0,\bm{s}}}{W_{j\bm{s}}}
\right).
\label{eq:rdmm-t-normal-path-marginal}
\end{equation}
Integrating out the shared precision variable gives the pathway-level multivariate Student-$t$ distribution
\begin{equation}
\bm{y}_j
\mid
\bm{S}_j=\bm{s}
\sim
\operatorname{t}_{p}\!\left(
\nu_{\bm{s}},
\bm{\mu}_{\bm{s}},
\bm{\Sigma}_{0,\bm{s}}
\right).
\label{eq:rdmm-t-path-marginal}
\end{equation}

Under the scale-matrix parameterisation adopted here, the multivariate Student-$t$ density is
\begin{equation}
t_p(\bm{y};\nu,\bm{\mu},\bm{\Sigma})
=
\frac{
\Gamma\!\left((\nu+p)/2\right)
}{
\Gamma(\nu/2)(\nu\pi)^{p/2}|\bm{\Sigma}|^{1/2}
}
\left[
1+
\frac{1}{\nu}
(\bm{y}-\bm{\mu})^{\top}
\bm{\Sigma}^{-1}
(\bm{y}-\bm{\mu})
\right]^{-(\nu+p)/2}.
\label{eq:rdmm-t-density}
\end{equation}
Here, $\bm{\Sigma}$ denotes the $t$ scale matrix rather than the covariance matrix. For $\nu>2$,
\begin{equation}
\operatorname{Cov}(\bm{y})
=
\frac{\nu}{\nu-2}\bm{\Sigma}.
\label{eq:rdmm-t-covariance}
\end{equation}

Let $\bm{\Theta}$ denote the collection of all model parameters. Using the factorised pathway probabilities in \eqref{eq:rdmm-t-pathprob}, the observed-data mixture density is
\begin{equation}
f(\bm{y}_j;\bm{\Theta})
=
\sum_{\bm{s}\in\mathcal{S}}
\pi_{\bm{s}}
t_p\!\left(
\bm{y}_j;
\nu_{\bm{s}},
\bm{\mu}_{\bm{s}},
\bm{\Sigma}_{0,\bm{s}}
\right),
\label{eq:rdmm-t-mixture}
\end{equation}
where $\pi_{\bm{s}}=\prod_{l=1}^{h}\pi_{s_l}^{(l)}$. The observed-data log-likelihood is therefore
\begin{equation}
\ell(\bm{\Theta})
=
\sum_{j=1}^{n}
\log\!\left[
\sum_{\bm{s}\in\mathcal{S}}
\pi_{\bm{s}}
t_p\!\left(
\bm{y}_j;
\nu_{\bm{s}},
\bm{\mu}_{\bm{s}},
\bm{\Sigma}_{0,\bm{s}}
\right)
\right].
\label{eq:rdmm-t-observed-loglik}
\end{equation}

\subsection{Posterior pathway quantities and robustness}
\label{sec:rdmm-t-properties}

For pathway $\bm{s}$, define the squared Mahalanobis distance
\begin{equation}
d_{j\bm{s}}^{2}
=
(\bm{y}_j-\bm{\mu}_{\bm{s}})^{\top}
\bm{\Sigma}_{0,\bm{s}}^{-1}
(\bm{y}_j-\bm{\mu}_{\bm{s}}).
\label{eq:rdmm-t-mahalanobis}
\end{equation}
The posterior probability that observation $j$ follows pathway $\bm{s}$ is
\begin{equation}
\tau_{j\bm{s}}
=
\Pr\!\left(\bm{S}_j=\bm{s}\mid\bm{y}_j\right)
=
\frac{
\pi_{\bm{s}}
t_p\!\left(
\bm{y}_j;
\nu_{\bm{s}},
\bm{\mu}_{\bm{s}},
\bm{\Sigma}_{0,\bm{s}}
\right)
}{
\displaystyle
\sum_{\bm{r}\in\mathcal{S}}
\pi_{\bm{r}}
t_p\!\left(
\bm{y}_j;
\nu_{\bm{r}},
\bm{\mu}_{\bm{r}},
\bm{\Sigma}_{0,\bm{r}}
\right)
}.
\label{eq:rdmm-t-path-posterior}
\end{equation}

Conditional on $\bm{y}_j$ and pathway $\bm{s}$, the shared precision variable has posterior distribution
\begin{equation}
W_{j\bm{s}}
\mid
\bm{y}_j,
\bm{S}_j=\bm{s}
\sim
\operatorname{gamma}\!\left(
\frac{\nu_{\bm{s}}+p}{2},
\frac{\nu_{\bm{s}}+d_{j\bm{s}}^{2}}{2}
\right).
\label{eq:rdmm-t-W-posterior}
\end{equation}
Consequently,
\begin{equation}
\widehat{w}_{j\bm{s}}
=
\mathbb{E}\!\left[
W_{j\bm{s}}
\mid
\bm{y}_j,
\bm{S}_j=\bm{s}
\right]
=
\frac{\nu_{\bm{s}}+p}{\nu_{\bm{s}}+d_{j\bm{s}}^{2}}.
\label{eq:rdmm-t-w-hat}
\end{equation}

When $\bm{y}_j$ lies far from the pathway centre $\bm{\mu}_{\bm{s}}$, the Mahalanobis distance $d_{j\bm{s}}^{2}$ is large and $\widehat{w}_{j\bm{s}}$ becomes small. Because the same precision variable scales every conditional covariance matrix along the pathway, a small posterior value of $W_{j\bm{s}}$ inflates the effective residual variation at all layers simultaneously. The observation is therefore down-weighted throughout the entire latent hierarchy rather than only in the observation equation.

For subsequent parameter estimation, the layer-specific posterior responsibility is obtained by marginalising the complete-pathway responsibilities:
\begin{equation}
\tau_{ja}^{(l)}
=
\Pr\!\left(S_j^{(l)}=a\mid\bm{y}_j\right)
=
\sum_{\bm{s}\in\mathcal{S}:\,s_l=a}
\tau_{j\bm{s}}.
\label{eq:rdmm-t-layer-posterior}
\end{equation}
Thus, although the layer indicators are independent a priori, their posterior probabilities are induced by the joint complete-pathway posterior.

\subsection{Special cases and identifiability}
\label{sec:rdmm-t-special-identifiability}

As $\nu_{\bm{s}}\to\infty$, the gamma mixing distribution degenerates at $W_{j\bm{s}}=1$, and the pathway-level Student-$t$ distribution converges to
\[
\mathcal{N}_{p}\!\left(
\bm{\mu}_{\bm{s}},
\bm{\Sigma}_{0,\bm{s}}
\right).
\]
The proposed model therefore reduces to the standard DGMM with independent layer-specific mixing proportions as a limiting special case. When $h=1$, the model reduces to a mixture of $t$ factor analysers under the corresponding factor-analytic covariance restrictions.

To reduce rotational indeterminacy and improve identifiability, we impose the standard deep factor-analytic dimension constraint
\[
p>r_1>\cdots>r_h\geq 1.
\]
At each layer $l$ and for every local component $a$, the matrix
\begin{equation}
(\bm{\Lambda}_a^{(l)})^{\top}
(\bm{\Psi}_a^{(l)})^{-1}
\bm{\Lambda}_a^{(l)}
\label{eq:rdmm-t-identifiability-matrix}
\end{equation}
is constrained to be diagonal, with positive diagonal entries arranged in strictly decreasing order. In practice, $\bm{\Psi}_a^{(l)}$ is often further restricted to be diagonal, which substantially reduces the number of free covariance parameters.

As in all finite mixture models, component labels are identifiable only up to permutation. After convergence, the local components may be ordered separately within each layer, for example by decreasing $\pi_a^{(l)}$. Complete pathways may then be reported in lexicographic order according to their layer-specific component indices, or in decreasing order of posterior pathway mass.

\section{Parameter estimation}
\label{sec:rdmm-t-estimation}

The parameters of the proposed RDMM are estimated by a layer-wise stochastic expectation--maximisation procedure. The discrete pathway indicators are Rao--Blackwellised by retaining their posterior responsibilities, whereas the shared precision variables and continuous latent vectors are sampled sequentially from their conditional distributions. With one Monte Carlo draw per observation and pathway, the procedure is a stochastic EM (SEM) algorithm~\citep{nielsen2000stochastic}; using multiple draws yields a more stable Monte Carlo EM (MCEM) approximation~\citep{levine2001implementations}.

Let
\[
\bm{\Theta}
=
\left\{
\pi_a^{(l)},
\bm{\eta}_a^{(l)},
\bm{\Lambda}_a^{(l)},
\bm{\Psi}_a^{(l)},
\nu_{\bm{s}}
:
 l=1,\ldots,h,\;
 a=1,\ldots,K_l,\;
 \bm{s}\in\mathcal{S}
\right\}
\]
denote the collection of model parameters. The independent layer-specific mixing proportions satisfy
\[
\pi_a^{(l)}>0,
\qquad
\sum_{a=1}^{K_l}\pi_a^{(l)}=1,
\qquad
l=1,\ldots,h.
\]

\subsection{Complete-data representation}
\label{sec:rdmm-t-complete-data}

For observation $j$ and pathway $\bm{s}\in\mathcal{S}$, define
\[
u_{j\bm{s}}
=
\mathbb{I}\!\left(\bm{S}_j=\bm{s}\right),
\qquad
\sum_{\bm{s}\in\mathcal{S}}u_{j\bm{s}}=1.
\]
Conditional on $u_{j\bm{s}}=1$, the missing data consist of the shared precision variable $W_{j\bm{s}}$ and the latent vectors
$\bm{z}_j^{(1)},\ldots,\bm{z}_j^{(h)}$. Using
$\bm{z}_j^{(0)}=\bm{y}_j$ and
$\pi_{\bm{s}}=\prod_{l=1}^{h}\pi_{s_l}^{(l)}$, the complete-data
log-likelihood is
\begin{align}
\ell_c(\bm{\Theta})
=
\sum_{j=1}^{n}
\sum_{\bm{s}\in\mathcal{S}}
u_{j\bm{s}}
\Bigg\{
&
\sum_{l=1}^{h}\log \pi_{s_l}^{(l)}
+
\log f_{\mathrm G}\!\left(
W_{j\bm{s}};
\frac{\nu_{\bm{s}}}{2},
\frac{\nu_{\bm{s}}}{2}
\right)
\nonumber\\
&+
\log \phi_{r_h}\!\left(
\bm{z}_j^{(h)};
\bm{0},
\frac{\bm{I}_{r_h}}{W_{j\bm{s}}}
\right)
\nonumber\\
&+
\sum_{l=1}^{h}
\log \phi_{r_{l-1}}\!\left(
\bm{z}_j^{(l-1)};
\bm{\eta}_{s_l}^{(l)}
+
\bm{\Lambda}_{s_l}^{(l)}\bm{z}_j^{(l)},
\frac{\bm{\Psi}_{s_l}^{(l)}}{W_{j\bm{s}}}
\right)
\Bigg\},
\label{eq:rdmm-sem-complete-loglik}
\end{align}
where $f_{\mathrm G}(\cdot;a,b)$ is the gamma density under the shape--rate parameterisation and $\phi_q(\cdot;\bm{\mu},\bm{\Sigma})$ is the $q$-dimensional Gaussian density. The deepest-layer Gaussian term contains no unknown location or covariance parameters under the adopted standardisation.

\subsection{Stochastic E-step}
\label{sec:rdmm-t-stochastic-estep}

Suppose that $\bm{\Theta}^{(t)}$ is available at iteration $t$. The stochastic E-step has three parts: calculation of the pathway responsibilities, generation of the shared precision variables, and sequential generation of the latent vectors from the observation layer to the deepest layer.

\subsubsection{Pathway responsibilities}

For each $\bm{s}\in\mathcal{S}$, define
\begin{equation}
\tau_{j\bm{s}}^{(t)}
=
\Pr\!\left(
\bm{S}_j=\bm{s}
\mid
\bm{y}_j;
\bm{\Theta}^{(t)}
\right)
=
\frac{
\pi_{\bm{s}}^{(t)}
 t_p\!\left(
 \bm{y}_j;
 \nu_{\bm{s}}^{(t)},
 \bm{\mu}_{\bm{s}}^{(t)},
 \bm{\Sigma}_{0,\bm{s}}^{(t)}
 \right)
}{
\displaystyle
\sum_{\bm{r}\in\mathcal{S}}
\pi_{\bm{r}}^{(t)}
 t_p\!\left(
 \bm{y}_j;
 \nu_{\bm{r}}^{(t)},
 \bm{\mu}_{\bm{r}}^{(t)},
 \bm{\Sigma}_{0,\bm{r}}^{(t)}
 \right)
},
\label{eq:rdmm-sem-tau}
\end{equation}
where
\[
\pi_{\bm{s}}^{(t)}
=
\prod_{l=1}^{h}\pi_{s_l}^{(l,t)}.
\]
For numerical stability, let
\[
A_{j\bm{s}}^{(t)}
=
\log \pi_{\bm{s}}^{(t)}
+
\log t_p\!\left(
\bm{y}_j;
\nu_{\bm{s}}^{(t)},
\bm{\mu}_{\bm{s}}^{(t)},
\bm{\Sigma}_{0,\bm{s}}^{(t)}
\right),
\]
and evaluate
\begin{equation}
\tau_{j\bm{s}}^{(t)}
=
\exp\!\left\{
A_{j\bm{s}}^{(t)}
-
\operatorname{LSE}_{\bm{r}\in\mathcal{S}}
A_{j\bm{r}}^{(t)}
\right\},
\label{eq:rdmm-sem-logsumexp}
\end{equation}
where $\operatorname{LSE}$ denotes the log-sum-exp operator. Unlike a fully simulated SEM allocation step, the responsibilities in \eqref{eq:rdmm-sem-tau} are retained as soft weights. 

\subsubsection{Shared precision variables}

For pathway $\bm{s}$, define
\begin{equation}
d_{j\bm{s}}^{2(t)}
=
\left(
\bm{y}_j-\bm{\mu}_{\bm{s}}^{(t)}
\right)^{\top}
\left(
\bm{\Sigma}_{0,\bm{s}}^{(t)}
\right)^{-1}
\left(
\bm{y}_j-\bm{\mu}_{\bm{s}}^{(t)}
\right).
\label{eq:rdmm-sem-mahalanobis}
\end{equation}
The conditional distribution of the shared precision is
\begin{equation}
W_{j\bm{s}}
\mid
\bm{y}_j,
\bm{S}_j=\bm{s};
\bm{\Theta}^{(t)}
\sim
\operatorname{gamma}\!\left(
\frac{\nu_{\bm{s}}^{(t)}+p}{2},
\frac{\nu_{\bm{s}}^{(t)}+d_{j\bm{s}}^{2(t)}}{2}
\right).
\label{eq:rdmm-sem-W-posterior}
\end{equation}
For $m=1,\ldots,M$, generate
\[
W_{j\bm{s},m}^{(t)}
\sim
p\!\left(
W_{j\bm{s}}
\mid
\bm{y}_j,
\bm{S}_j=\bm{s};
\bm{\Theta}^{(t)}
\right).
\]
The exact conditional moments
\begin{align}
\widehat{w}_{j\bm{s}}^{(t)}
&=
\mathbb{E}\!\left(
W_{j\bm{s}}
\mid
\bm{y}_j,
\bm{S}_j=\bm{s};
\bm{\Theta}^{(t)}
\right)
=
\frac{\nu_{\bm{s}}^{(t)}+p}
{\nu_{\bm{s}}^{(t)}+d_{j\bm{s}}^{2(t)}},
\label{eq:rdmm-sem-EW}
\\
\widehat{e}_{j\bm{s}}^{(t)}
&=
\mathbb{E}\!\left(
\log W_{j\bm{s}}
\mid
\bm{y}_j,
\bm{S}_j=\bm{s};
\bm{\Theta}^{(t)}
\right)
\nonumber\\
&=
\psi\!\left(
\frac{\nu_{\bm{s}}^{(t)}+p}{2}
\right)
-
\log\!\left(
\frac{\nu_{\bm{s}}^{(t)}+d_{j\bm{s}}^{2(t)}}{2}
\right)
\label{eq:rdmm-sem-ElogW}
\end{align}
are used in the degrees-of-freedom update to avoid unnecessary Monte Carlo noise.

\subsubsection{Sequential generation of the latent vectors}

For a fixed pathway $\bm{s}$, the unit-scale marginal distribution of the layer-$l$ latent vector is
\[
\bm{z}_j^{(l)}
\mid
\bm{S}_j=\bm{s},
W_{j\bm{s}}
\sim
\mathcal{N}_{r_l}\!\left(
\bm{\mu}_{\bm{s}}^{(l)},
\frac{\bm{\Sigma}_{\bm{s}}^{(l)}}{W_{j\bm{s}}}
\right),
\]
where $\bm{\mu}_{\bm{s}}^{(l)}$ and $\bm{\Sigma}_{\bm{s}}^{(l)}$ are obtained from the recursions in \eqref{eq:rdmm-t-mu-recursion}--\eqref{eq:rdmm-t-sigma-recursion}. Combining this marginal distribution with the layer-$l$ transition gives
\begin{equation}
\bm{z}_j^{(l)}
\mid
\bm{z}_j^{(l-1)},
\bm{S}_j=\bm{s},
W_{j\bm{s}};
\bm{\Theta}^{(t)}
\sim
\mathcal{N}_{r_l}\!\left(
\bm{\rho}_{j\bm{s}}^{(l,t)},
\frac{\bm{\Omega}_{\bm{s}}^{(l,t)}}{W_{j\bm{s}}}
\right),
\label{eq:rdmm-sem-latent-conditional}
\end{equation}
where
\begin{align}
\bm{\Omega}_{\bm{s}}^{(l,t)}
&=
\Bigg[
\left(\bm{\Sigma}_{\bm{s}}^{(l,t)}\right)^{-1}
+
\left(\bm{\Lambda}_{s_l}^{(l,t)}\right)^{\top}
\left(\bm{\Psi}_{s_l}^{(l,t)}\right)^{-1}
\bm{\Lambda}_{s_l}^{(l,t)}
\Bigg]^{-1},
\label{eq:rdmm-sem-Xi}
\\
\bm{\rho}_{j\bm{s}}^{(l,t)}
&=
\bm{\Omega}_{\bm{s}}^{(l,t)}
\Bigg[
\left(\bm{\Sigma}_{\bm{s}}^{(l,t)}\right)^{-1}
\bm{\mu}_{\bm{s}}^{(l,t)}
\nonumber\\
&\hspace{3.0cm}
+
\left(\bm{\Lambda}_{s_l}^{(l,t)}\right)^{\top}
\left(\bm{\Psi}_{s_l}^{(l,t)}\right)^{-1}
\left
\{
\bm{z}_j^{(l-1)}
-
\bm{\eta}_{s_l}^{(l,t)}
\right\}
\Bigg].
\label{eq:rdmm-sem-rho}
\end{align}
The conditional mean in \eqref{eq:rdmm-sem-rho} does not depend on the value of $W_{j\bm{s}}$, whereas the conditional covariance is divided by the same precision variable shared across all layers.

For every $j$, $\bm{s}$, and Monte Carlo replicate $m$, set
\[
\bm{z}_{j\bm{s},m}^{(0,t)}=\bm{y}_j
\]
and generate sequentially, for $l=1,\ldots,h$,
\begin{equation}
\bm{z}_{j\bm{s},m}^{(l,t)}
\sim
\mathcal{N}_{r_l}\!\left(
\bm{\rho}_{j\bm{s},m}^{(l,t)},
\frac{\bm{\Omega}_{\bm{s}}^{(l,t)}}
{W_{j\bm{s},m}^{(t)}}
\right),
\label{eq:rdmm-sem-sequential-draw}
\end{equation}
where $\bm{\rho}_{j\bm{s},m}^{(l,t)}$ is obtained from \eqref{eq:rdmm-sem-rho} by replacing $\bm{z}_j^{(l-1)}$ with the current draw $\bm{z}_{j\bm{s},m}^{(l-1,t)}$. This forward stochastic recursion is the robust counterpart of the latent-variable generation step in the DGMM. The essential difference is that the same draw $W_{j\bm{s},m}^{(t)}$ scales the conditional covariance at every layer of the pathway.

\subsection{Monte Carlo sufficient statistics}
\label{sec:rdmm-t-mc-statistics}

Define the augmented latent regressor and the corresponding coefficient matrix by
\[
\widetilde{\bm{z}}_{j\bm{s},m}^{(l,t)}
=
\begin{pmatrix}
1\\
\bm{z}_{j\bm{s},m}^{(l,t)}
\end{pmatrix},
\qquad
\bm{B}_a^{(l)}
=
\begin{pmatrix}
\bm{\eta}_a^{(l)} & \bm{\Lambda}_a^{(l)}
\end{pmatrix}.
\]
For each observation, pathway, and layer, define
\begin{align}
\widehat{\bm{R}}_{j\bm{s}}^{(l,t)}
&=
\frac{1}{M}
\sum_{m=1}^{M}
W_{j\bm{s},m}^{(t)}
\bm{z}_{j\bm{s},m}^{(l-1,t)}
\widetilde{\bm{z}}_{j\bm{s},m}^{(l,t)\top},
\label{eq:rdmm-sem-Rjs}
\\
\widehat{\bm{Q}}_{j\bm{s}}^{(l,t)}
&=
\frac{1}{M}
\sum_{m=1}^{M}
W_{j\bm{s},m}^{(t)}
\widetilde{\bm{z}}_{j\bm{s},m}^{(l,t)}
\widetilde{\bm{z}}_{j\bm{s},m}^{(l,t)\top}.
\label{eq:rdmm-sem-Qjs}
\end{align}
For a candidate coefficient matrix $\bm{B}$, also define
\begin{align}
\widehat{\bm{H}}_{j\bm{s}}^{(l,t)}(\bm{B})
=
\frac{1}{M}
\sum_{m=1}^{M}
&W_{j\bm{s},m}^{(t)}
\left(
\bm{z}_{j\bm{s},m}^{(l-1,t)}
-
\bm{B}\widetilde{\bm{z}}_{j\bm{s},m}^{(l,t)}
\right)
\nonumber\\
&\times
\left(
\bm{z}_{j\bm{s},m}^{(l-1,t)}
-
\bm{B}\widetilde{\bm{z}}_{j\bm{s},m}^{(l,t)}
\right)^{\top}.
\label{eq:rdmm-sem-Hjs}
\end{align}
These quantities approximate the weighted complete-data moments required in the M-step. The factor $W_{j\bm{s},m}^{(t)}$ is what distinguishes the robust updates from their Gaussian DGMM counterparts.

\subsection{M-step}
\label{sec:rdmm-t-mstep}

\subsubsection{Layer-specific mixing proportions}

Because the component indicators are independent a priori across layers, the mixing proportions are updated locally rather than as unrestricted pathway probabilities. For $l=1,\ldots,h$ and $a=1,\ldots,K_l$, define
\[
N_a^{(l,t)}
=
\sum_{j=1}^{n}
\sum_{\bm{s}\in\mathcal{S}:\,s_l=a}
\tau_{j\bm{s}}^{(t)}.
\]
Then
\begin{equation}
\pi_a^{(l,t+1)}
=
\frac{N_a^{(l,t)}}{n}
=
\frac{1}{n}
\sum_{j=1}^{n}
\sum_{\bm{s}\in\mathcal{S}:\,s_l=a}
\tau_{j\bm{s}}^{(t)}.
\label{eq:rdmm-sem-pi-update}
\end{equation}
This update preserves $\pi_{\bm{s}}=\prod_{l=1}^{h}\pi_{s_l}^{(l)}$ and does not introduce parent-dependent transition probabilities.

\subsubsection{Location vectors and loading matrices}

For local component $a$ at layer $l$, aggregate the Monte Carlo sufficient statistics over all pathways containing that component:
\begin{align}
\bm{R}_a^{(l,t)}
&=
\sum_{j=1}^{n}
\sum_{\bm{s}\in\mathcal{S}:\,s_l=a}
\tau_{j\bm{s}}^{(t)}
\widehat{\bm{R}}_{j\bm{s}}^{(l,t)},
\label{eq:rdmm-sem-Ra}
\\
\bm{Q}_a^{(l,t)}
&=
\sum_{j=1}^{n}
\sum_{\bm{s}\in\mathcal{S}:\,s_l=a}
\tau_{j\bm{s}}^{(t)}
\widehat{\bm{Q}}_{j\bm{s}}^{(l,t)}.
\label{eq:rdmm-sem-Qa}
\end{align}
The weighted least-squares update is
\begin{equation}
\bm{B}_a^{(l,t+1)}
=
\bm{R}_a^{(l,t)}
\left(\bm{Q}_a^{(l,t)}\right)^{-1}.
\label{eq:rdmm-sem-B-update}
\end{equation}
The first column of $\bm{B}_a^{(l,t+1)}$ gives $\bm{\eta}_a^{(l,t+1)}$, and its remaining columns give $\bm{\Lambda}_a^{(l,t+1)}$. In implementation, a generalised inverse or a small ridge term may be used if $\bm{Q}_a^{(l,t)}$ is nearly singular.

\subsubsection{Specific covariance matrices}

The unrestricted covariance update is
\begin{equation}
\bm{\Psi}_a^{(l,t+1)}
=
\frac{1}{N_a^{(l,t)}}
\sum_{j=1}^{n}
\sum_{\bm{s}\in\mathcal{S}:\,s_l=a}
\tau_{j\bm{s}}^{(t)}
\widehat{\bm{H}}_{j\bm{s}}^{(l,t)}
\!\left(\bm{B}_a^{(l,t+1)}\right).
\label{eq:rdmm-sem-Psi-update-full}
\end{equation}
If a diagonal-specific covariance is imposed, use
\begin{equation}
\bm{\Psi}_a^{(l,t+1)}
=
\operatorname{diag}\!\left[
\frac{1}{N_a^{(l,t)}}
\sum_{j=1}^{n}
\sum_{\bm{s}\in\mathcal{S}:\,s_l=a}
\tau_{j\bm{s}}^{(t)}
\widehat{\bm{H}}_{j\bm{s}}^{(l,t)}
\!\left(\bm{B}_a^{(l,t+1)}\right)
\right].
\label{eq:rdmm-sem-Psi-update-diag}
\end{equation}
The denominator is the effective component count $N_a^{(l,t)}$, rather than a sum of precision weights, because the complete-data Gaussian log-determinant term is repeated once for each effective observation. To avoid degeneracy, the eigenvalues, or the diagonal entries under a diagonal restriction, may be bounded below by a small constant $\psi_{\min}>0$.

\subsection{Degrees-of-freedom update}
\label{sec:rdmm-t-nu-update}

Let
\[
N_{\bm{s}}^{(t)}
=
\sum_{j=1}^{n}\tau_{j\bm{s}}^{(t)}.
\]
For a pathway-specific degrees-of-freedom parameter, update
$\nu_{\bm{s}}$ by solving
\begin{equation}
\log\!\left(\frac{\nu_{\bm{s}}}{2}\right)
-
\psi\!\left(\frac{\nu_{\bm{s}}}{2}\right)
+1
+
\frac{1}{N_{\bm{s}}^{(t)}}
\sum_{j=1}^{n}
\tau_{j\bm{s}}^{(t)}
\left(
\widehat{e}_{j\bm{s}}^{(t)}
-
\widehat{w}_{j\bm{s}}^{(t)}
\right)
=0.
\label{eq:rdmm-sem-nu-equation}
\end{equation}
This one-dimensional equation can be solved by a safeguarded Newton method or bisection over
\[
2+\epsilon
\leq
\nu_{\bm{s}}
\leq
\nu_{\max}.
\]
The lower bound ensures a finite pathway covariance, whereas a sufficiently large $\nu_{\max}$ approximates the Gaussian limit. If some pathways have small effective counts, the degrees of freedom may be shared across related pathways, for example
\[
\nu_{\bm{s}}=\nu_{s_1},
\]
in which case the terms in \eqref{eq:rdmm-sem-nu-equation} are pooled across all pathways that share the corresponding first-layer component.

\subsection{Stochastic estimation algorithm}
\label{sec:rdmm-t-sem-algorithm}

The complete estimation procedure is summarised as follows.
\begin{enumerate}
\item Initialise $\bm{\Theta}^{(0)}$ and choose the number $M$ of Monte Carlo draws. Setting $M=1$ gives the SEM version; $M>1$ gives a stabilised MCEM version.

\item At iteration $t$, compute
      $\bm{\mu}_{\bm{s}}^{(l,t)}$ and
      $\bm{\Sigma}_{\bm{s}}^{(l,t)}$ for every pathway and layer using the
      recursive formulas in
      \eqref{eq:rdmm-t-mu-recursion}--\eqref{eq:rdmm-t-sigma-recursion}.

\item Compute the pathway responsibilities
      $\tau_{j\bm{s}}^{(t)}$ using
      \eqref{eq:rdmm-sem-tau} or its log-sum-exp implementation
      \eqref{eq:rdmm-sem-logsumexp}.

\item For every observation $j$, pathway $\bm{s}$, and replicate
      $m=1,\ldots,M$, draw $W_{j\bm{s},m}^{(t)}$ from
      \eqref{eq:rdmm-sem-W-posterior}. Conditional on that same precision draw,
      generate
      $\bm{z}_{j\bm{s},m}^{(1,t)},\ldots,
       \bm{z}_{j\bm{s},m}^{(h,t)}$
      sequentially using \eqref{eq:rdmm-sem-sequential-draw}.

\item Form the Monte Carlo sufficient statistics in
      \eqref{eq:rdmm-sem-Rjs}--\eqref{eq:rdmm-sem-Hjs}.

\item Update the independent layer-specific mixing proportions with
      \eqref{eq:rdmm-sem-pi-update}.

\item For $l=1,\ldots,h$ and every local component $a$, update
      $\bm{\eta}_a^{(l)}$, $\bm{\Lambda}_a^{(l)}$, and
      $\bm{\Psi}_a^{(l)}$ using
      \eqref{eq:rdmm-sem-B-update} and either
      \eqref{eq:rdmm-sem-Psi-update-full} or
      \eqref{eq:rdmm-sem-Psi-update-diag}.

\item Update the degrees-of-freedom parameters by solving
      \eqref{eq:rdmm-sem-nu-equation}.

\item Reimpose the identifiability constraints and covariance lower bounds,
      evaluate the observed-data log-likelihood
      \eqref{eq:rdmm-t-observed-loglik}, and iterate.
\end{enumerate}

This algorithm retains the essential layer-wise structure of the DGMM estimation procedure. In the Gaussian limit $\nu_{\bm{s}}\rightarrow\infty$, the shared precision converges to one, and the weighted updates reduce to their DGMM counterparts.

\subsection{Convergence and practical implementation}
\label{sec:rdmm-t-sem-practical}

Because the stochastic step introduces Monte Carlo variability, the observed-data log-likelihood is not required to increase at every SEM iteration. For $M=1$, a burn-in period of $T_{\mathrm{burn}}$ iterations may be discarded, after which parameter iterates can be retained and averaged after component labels have been aligned and the identifiability constraints have been imposed. Alternatively, the retained iterate with the largest observed log-likelihood may be selected. For the MCEM version, $M$ may be increased as the algorithm approaches convergence, and a moving average of the log-likelihood can be monitored.

Initial values can be obtained from a fitted DGMM or from a layer-wise Gaussian mixture/factor-analysis initialisation. The loading matrices may be initialised using principal component directions, the specific covariance matrices may be initialised as diagonal positive matrices, and the degrees-of-freedom parameters may be started at moderate values such as $\nu_{\bm{s}}^{(0)}\in[4,30].$ Multiple random starts are recommended. Pathways with persistently negligible effective counts may be pruned, or their degrees-of-freedom parameters may be pooled with those of related pathways.

After convergence, observation $j$ may be assigned to the complete pathway
\[
\widehat{\bm{s}}_j
=
\arg\max_{\bm{s}\in\mathcal{S}}
\tau_{j\bm{s}},
\]
or, when only the first-layer cluster is of interest, to
\begin{equation}
\widehat{s}_j^{(1)}
=
\arg\max_{a=1,\ldots,K_1}
\sum_{\bm{s}\in\mathcal{S}:\,s_1=a}
\tau_{j\bm{s}}.
\label{eq:rdmm-sem-first-layer-classification}
\end{equation}

\section{Numerical simulations}
\label{sec:rdmm-t-simulation}
\subsection{Data-generating model}
\label{sec:simulation-model}

The same two-layer Student-$t$ deep mixture model was used throughout both simulation studies. The observed dimension was $p=20$, the latent dimensions were $(r_1,r_2)=(5,2)$, and each latent layer contained two mixture components, $(K_1,K_2)=(2,2)$. Consequently, the complete pathway set was $\mathcal S=\{1,2\}\times\{1,2\},$ with equal mixing proportions $ \bm\pi^{(1)}=\bm\pi^{(2)}=(0.5,0.5)^\top, $ so that each pathway had probability $0.25$.

Data were generated according to the hierarchical Student-$t$ deep mixture model described in Section~\ref{sec:rdmm-t-model}. Specifically, for observation $j$, a latent precision variable was generated from
\[
W_j\sim
\operatorname{gamma}\!\left(\frac{\nu}{2},\frac{\nu}{2}\right),
\]
where the gamma distribution is parameterized by shape and rate. The same realization of $W_j$ was shared across both latent layers and the observation model, inducing observation-level heavy-tailed behavior throughout the entire hierarchy.

The component parameters were fixed throughout the simulation studies. The first-layer locations were
\[
\bm\eta_1^{(1)}
=
-2.5\sum_{q=1}^{3}\bm e_q^{(20)},
\qquad
\bm\eta_2^{(1)}
=
2.5\sum_{q=1}^{3}\bm e_q^{(20)},
\]
while the second-layer locations were
$
\bm\eta_1^{(2)}
=
1.25\bm e_1^{(5)},
\qquad
\bm\eta_2^{(2)}
=
-1.25\bm e_1^{(5)}.
$

Let $\bm a=(1.0,0.9,1.1,0.8)^\top, \bm A=\bm I_5\otimes\bm a, $ and $\bm D_2=\operatorname{diag}(1.10,0.90,1.05,0.95,1.00).$ The first-layer loading matrices were specified as $\bm\Lambda_1^{(1)}= \bm A, \bm\Lambda_2^{(1)}=\bm A\bm D_2.$ The second-layer loading matrices were
\[
\bm\Lambda_1^{(2)}
=
\begin{pmatrix}
1.0&0.0\\
0.8&0.2\\
0.0&1.0\\
0.2&0.8\\
0.6&-0.6
\end{pmatrix},
\qquad
\bm\Lambda_2^{(2)}
=
\begin{pmatrix}
0.9&0.1\\
0.7&-0.2\\
0.1&0.9\\
-0.2&0.7\\
0.5&0.5
\end{pmatrix}.
\]
Residual covariance matrices were diagonal and common within each layer, $\bm\Psi_a^{(1)}=0.5\bm I_{20},a=1,2,$ and $\bm\Psi_b^{(2)}=0.3\bm I_5,b=1,2.$ These parameter values were identical in both simulation studies. Simulation~1 varied the common degrees of freedom to investigate the effect of tail heaviness, whereas Simulation~2 fixed the clean-data distribution and introduced heavy-tailed observation-level contamination to assess robustness.

\subsection{Simulation 1: Parameter recovery}
\label{sec:study1}

Simulation~1 investigated the effect of tail heaviness on parameter recovery and clustering performance. No external contamination was introduced, and the common degrees of freedom used to generate the data varied over
$$
\nu\in\{3,3.5,4,4.5,5,5.5\}.
$$
The smaller values correspond to heavier tails, whereas larger values produce distributions closer to the Gaussian case. For each value of $\nu$, $n=1000$ observations were generated, and the experiment was repeated over 500 independent Monte Carlo replications.

RDMM was fitted under the common degrees-of-freedom specification, while DGMM was fitted using the same hierarchical architecture but without the observation-level latent precision. The numbers of latent layers, mixture components, and latent dimensions were fixed at their true values, $(K_1,K_2)=(2,2)$ and $(r_1,r_2)=(5,2)$, for both methods. Performance was evaluated at both the first-layer and pathway levels using the adjusted Rand index (ARI) and the misclassification rate (MR). In addition, the Monte Carlo mean and standard deviation of the estimated common degrees of freedom were recorded for RDMM.

Table~\ref{tab:study1-results} summarizes the recovery of the common degrees of freedom together with the clustering performance. Across all simulation settings, RDMM accurately recovered the true common degrees of freedom. The Monte Carlo mean estimates were consistently close to the generating values, with absolute biases below 0.06 in every scenario. Moreover, the estimated standard deviations remained small, indicating stable and reliable estimation even under pronounced heavy-tailed distributions.

\begin{table}[t]
\centering
\caption{Recovery of the common degrees of freedom and clustering performance in Simulation~1. Entries are Monte Carlo means, with standard deviations in parentheses.}
\label{tab:study1-results}
\begin{tabular}{ccccccc}
\hline
$\nu$ & Method & Estimated $\nu$ & First-layer ARI & First-layer MR & Pathway ARI & Pathway MR\\
\hline
3.0 & DGMM & -- & 0.167 (0.273) & 0.377 (0.164) & 0.074 (0.111) & 0.660 (0.091)\\
    & RDMM & 2.998 (0.151) & 0.860 (0.172) & 0.045 (0.088) & 0.671 (0.149) & 0.156 (0.127)\\
\hline
3.5 & DGMM & -- & 0.418 (0.347) & 0.236 (0.187) & 0.163 (0.132) & 0.583 (0.111)\\
    & RDMM & 3.488 (0.172) & 0.895 (0.129) & 0.031 (0.065) & 0.706 (0.115) & 0.133 (0.096)\\
\hline
4.0 & DGMM & -- & 0.626 (0.293) & 0.128 (0.135) & 0.259 (0.128) & 0.517 (0.094)\\
    & RDMM & 3.978 (0.192) & 0.923 (0.061) & 0.020 (0.029) & 0.738 (0.065) & 0.112 (0.051)\\
\hline
4.5 & DGMM & -- & 0.782 (0.214) & 0.066 (0.086) & 0.352 (0.118) & 0.464 (0.097)\\
    & RDMM & 4.476 (0.230) & 0.938 (0.016) & 0.016 (0.004) & 0.745 (0.060) & 0.110 (0.044)\\
\hline
5.0 & DGMM & -- & 0.868 (0.123) & 0.036 (0.042) & 0.424 (0.107) & 0.420 (0.114)\\
    & RDMM & 4.965 (0.271) & 0.947 (0.014) & 0.014 (0.004) & 0.760 (0.049) & 0.101 (0.031)\\
\hline
5.5 & DGMM & -- & 0.903 (0.075) & 0.025 (0.022) & 0.456 (0.094) & 0.403 (0.120)\\
    & RDMM & 5.446 (0.290) & 0.951 (0.013) & 0.012 (0.003) & 0.762 (0.059) & 0.101 (0.041)\\
\hline
\end{tabular}
\end{table}

Regarding clustering performance, both RDMM and DGMM exhibited gradual improvements as the true degrees of freedom increased, reflecting the diminishing influence of heavy tails. Nevertheless, RDMM consistently outperformed DGMM across the entire range of simulated values. At the first-layer level, the mean ARI of RDMM increased from 0.860 to 0.951, whereas that of DGMM improved from only 0.167 to 0.903. The superiority of RDMM was particularly pronounced under severe heavy-tailed settings. For example, when $\nu=3$, RDMM achieved a first-layer ARI of 0.860 compared with only 0.167 for DGMM, while the corresponding misclassification rates were 0.045 and 0.377, respectively. Similar improvements were observed for complete-pathway clustering, where RDMM consistently produced substantially higher ARI values and markedly lower misclassification rates than DGMM. Although the performance gap narrowed as the distributions became closer to Gaussian, RDMM remained uniformly superior across all settings.

\begin{figure}[!ht]
    \centering
    \includegraphics[width=\linewidth]{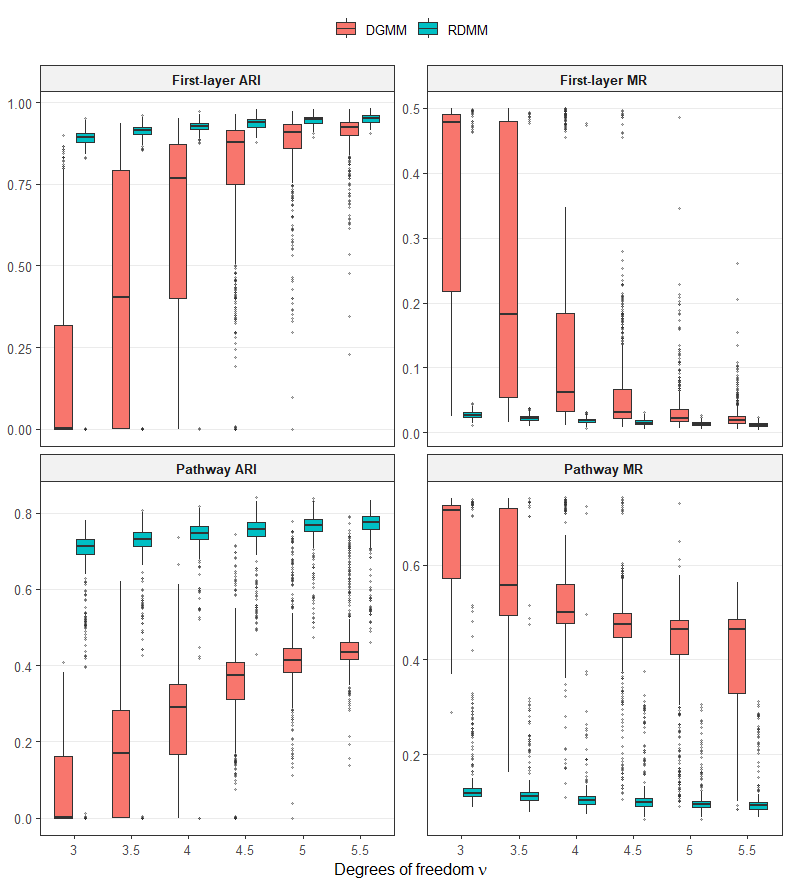}
\caption{
Clustering performance of DGMM and RDMM under varying degrees-of-freedom settings. Boxplots display the distributions of the ARI and MR over 500 Monte Carlo replications for both first-layer and complete-pathway clustering.
}
    \label{fig:study1-boxplots}
\end{figure}

Figure~\ref{fig:study1-boxplots} presents the replication-level distributions of the four clustering measures across different degrees-of-freedom settings. The graphical summaries are fully consistent with the numerical results reported in Table~\ref{tab:study1-results}. Overall, RDMM consistently achieved higher median ARI values and lower median misclassification rates than DGMM for both first-layer and complete-pathway clustering.

The superiority of RDMM was particularly evident under heavy-tailed settings ($\nu=3$ and $\nu=3.5$), where DGMM exhibited substantial performance deterioration, as reflected by low median ARI values, high misclassification rates, and considerable between-replication variability. In contrast, RDMM maintained high clustering accuracy with relatively concentrated boxplots, demonstrating strong robustness to heavy-tailed observations. As the degrees of freedom increased, the performance of DGMM gradually improved, resulting in higher median ARI values and lower misclassification rates. Nevertheless, RDMM consistently remained superior and generally exhibited narrower interquartile ranges, indicating greater stability across Monte Carlo replications. The advantage of RDMM was especially pronounced for complete-pathway clustering, where a substantial performance gap persisted even under relatively light-tailed distributions. Overall, these results demonstrate that introducing a shared latent scale variable substantially improves both the robustness and stability of deep mixture modelling when the Gaussian assumption is violated.

\subsection{Simulation 2: Heavy-tailed contamination}
\label{sec:study2}

Simulation~2 investigated the robustness of RDMM under increasing observation-level contamination. Each Monte Carlo replication began with a clean sample of size $n=1000$ generated from the same two-layer Student-$t$ deep mixture model described in Section~\ref{sec:simulation-model}. The clean data-generating model used a common degrees-of-freedom parameter fixed at $\nu_0=10.$ Thus, the uncontaminated observations were moderately heavy-tailed, but were considerably closer to Gaussian observations than those generated under the smaller values of $\nu$ examined in Simulation~1.

The contamination proportions were $\epsilon\in \{0,0.1,0.2,0.3,0.4,0.5\}$. For each replication, a random permutation $(o_1,\ldots,o_n)$ of the observation indices was generated. At contamination level $\epsilon$, the observations indexed by $\mathcal O_{\epsilon}=\{o_1,\ldots,o_{\lfloor\epsilon n\rfloor}\}$ were designated as contaminated observations.

For every observation $j$, an independent 20-dimensional heavy-tailed perturbation was generated according to $\bm\delta_j\sim t_{20}\!\left(\bm 0, 2\bm I_{20}, 4 \right)$, where $2\bm I_{20}$ is the scale matrix and the perturbation degrees of freedom are equal to four. Equivalently, the perturbation was generated using the scale-mixture representation
\[
\bm g_j\sim
\mathcal N_{20}\!\left(
\bm 0,
2\bm I_{20}
\right),
\qquad
U_j\sim\chi_4^2,
\]
independently, followed by
\[
\bm\delta_j
=
\frac{\bm g_j}{\sqrt{U_j/4}}.
\]
Consequently, $\operatorname{Cov}(\bm\delta_j)=4\bm I_{20}$, while the fourth moments of the perturbation distribution do not exist. The perturbations therefore contain occasional extreme observations without introducing a systematic directional location shift.

The contaminated observation at level $\epsilon$ was defined as
\[
\bm y_{j,\epsilon}^{\mathrm{cont}}
=
\begin{cases}
\bm y_j+\bm\delta_j,
&
j\in\mathcal O_{\epsilon},
\\[3pt]
\bm y_j,
&
j\notin\mathcal O_{\epsilon}.
\end{cases}
\]
The perturbation was applied to the complete 20-dimensional observation vector. Therefore, contamination occurred at the observation level rather than independently at individual coordinates. Within each replication, all contamination levels were constructed from the same clean sample, the same random ordering of observations, and the same set of perturbation vectors. The contaminated sets were therefore nested: $\mathcal O_{0.1}\subset\mathcal O_{0.2}\subset\mathcal O_{0.3}\subset\mathcal O_{0.4}\subset\mathcal O_{0.5}.$ This paired construction reduces extraneous Monte Carlo variation when comparing the effect of increasing contamination.

Although the clean observations were generated using the common value $\nu_0=10$, RDMM was fitted with four pathway-specific degrees-of-freedom parameters, $\nu_{11}, \nu_{12}, \nu_{21}, \nu_{22}$, one for each complete pathway. All four parameters were initialized at five and estimated from the contaminated data. This more flexible specification allows the fitted RDMM to adapt differently across pathways when contaminated observations are not uniformly allocated among the four latent subpopulations. Because the true clean model has a common degrees-of-freedom structure, the pathway-specific estimates in this simulation are interpreted as robustness diagnostics rather than as estimates of four distinct generating parameters. DGMM was fitted using the same two-layer architecture, numbers of mixture components, latent dimensions, and initialization strategy, but without the Student-$t$ precision variables. Both models were evaluated using the adjusted Rand index (ARI) and misclassification rate (MR) at the first-layer and complete-pathway levels. For each contamination condition, the experiment was repeated over $500$ independent Monte Carlo replications.

\begin{table}[!ht]
\centering
\caption{Clustering performance under heavy-tailed observation-level contamination in Simulation~2. Entries are Monte Carlo means, with standard deviations in parentheses. The mean estimated pathway-specific degrees of freedom ($\hat{\nu}$) is reported for RDMM only.}
\label{tab:study2-clustering}
\begin{tabular}{clccccc}
\hline
Contamination & Method & Estimated $\hat{\bar{\nu}}$ & First-layer ARI &
First-layer MR & Pathway ARI & Pathway MR\\
\hline
0.00 & DGMM & --            & 0.967 (0.019) & 0.008 (0.005) & 0.587 (0.127) & 0.250 (0.126)\\
     & RDMM & 10.212 (0.702)& 0.976 (0.010) & 0.006 (0.002) & 0.777 (0.080) & 0.099 (0.056)\\
\hline
0.10 & DGMM & --            & 0.683 (0.252) & 0.105 (0.121) & 0.381 (0.170) & 0.398 (0.162)\\
     & RDMM & 6.287 (0.621) & 0.934 (0.017) & 0.017 (0.004) & 0.712 (0.079) & 0.129 (0.059)\\
\hline
0.20 & DGMM & --            & 0.266 (0.331) & 0.324 (0.189) & 0.146 (0.179) & 0.591 (0.164)\\
     & RDMM & 4.451 (0.464) & 0.899 (0.020) & 0.026 (0.005) & 0.662 (0.082) & 0.154 (0.061)\\
\hline
0.30 & DGMM & --            & 0.134 (0.275) & 0.406 (0.158) & 0.074 (0.144) & 0.656 (0.138)\\
     & RDMM & 3.483 (0.226) & 0.863 (0.046) & 0.036 (0.022) & 0.615 (0.092) & 0.180 (0.072)\\
\hline
0.40 & DGMM & --            & 0.132 (0.266) & 0.407 (0.157) & 0.071 (0.137) & 0.654 (0.141)\\
     & RDMM & 3.047 (0.146) & 0.833 (0.027) & 0.044 (0.007) & 0.580 (0.087) & 0.199 (0.071)\\
\hline
0.50 & DGMM & --            & 0.134 (0.258) & 0.404 (0.157) & 0.069 (0.126) & 0.654 (0.132)\\
     & RDMM & 2.850 (0.126) & 0.800 (0.047) & 0.053 (0.021) & 0.536 (0.091) & 0.225 (0.076)\\
\hline
\end{tabular}
\end{table}

Table~\ref{tab:study2-clustering} summarizes the clustering performance together with the average estimated pathway-specific degrees of freedom under increasing levels of observation-level heavy-tailed contamination. As the contamination proportion increased, the clustering performance of both methods deteriorated; however, the degradation was substantially slower for RDMM. At the first-layer level, the mean ARI of RDMM decreased gradually from 0.976 under clean data to 0.800 at the highest contamination level, whereas the corresponding value for DGMM dropped sharply from 0.967 to only 0.134. Conversely, the first-layer misclassification rate of RDMM increased only moderately from 0.006 to 0.053, while that of DGMM increased dramatically from 0.008 to approximately 0.404. Similar trends were observed for complete-pathway clustering. The pathway ARI of RDMM decreased from 0.777 to 0.536, whereas that of DGMM declined from 0.587 to below 0.070. Likewise, the pathway misclassification rate of RDMM increased from 0.099 to 0.225, compared with an increase from 0.250 to approximately 0.654 for DGMM. These results demonstrate that RDMM provides substantially greater robustness against observation-level contamination than DGMM.

The estimated pathway-specific degrees of freedom further illustrate the adaptive behavior of RDMM. Although the uncontaminated data were generated from a common Student-$t$ distribution with $\nu_0=10$, the fitted model was allowed to estimate pathway-specific degrees of freedom. Under the uncontaminated setting, the average estimated value was 10.212, closely matching the generating parameter. As the contamination proportion increased, the average estimated degrees of freedom decreased monotonically from 10.212 to 2.850, indicating that RDMM automatically adapted by fitting progressively heavier-tailed component distributions. This adaptive reduction in the estimated degrees of freedom suggests that the latent scale variables effectively accommodated increasingly extreme observations, thereby preserving accurate and stable clustering performance even under substantial contamination.

\begin{figure}[!ht]
    \centering
    \includegraphics[width=\linewidth]{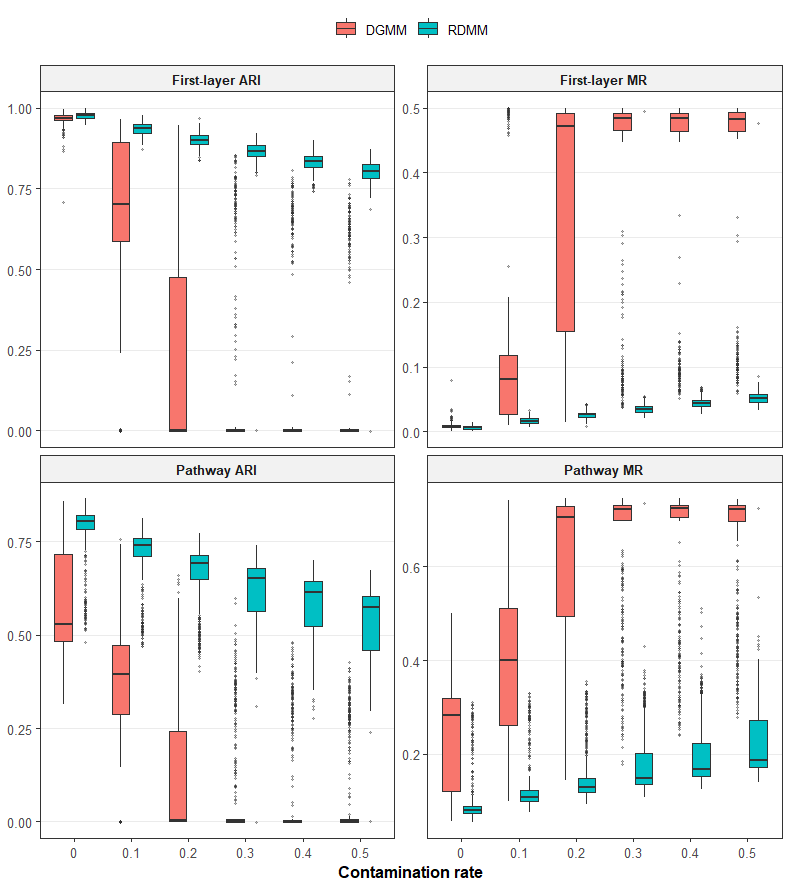}
\caption{
Effect of contamination on clustering performance for DGMM and RDMM. Boxplots show the ARI and MR over 500 Monte Carlo replications for first-layer and complete-pathway clustering at different contamination rates.
}
    \label{fig:study2-boxplots}
\end{figure}

Figure~\ref{fig:study2-boxplots} presents the replication-level distributions of the four clustering measures across different contamination rates. The graphical summaries are fully consistent with the numerical results reported in Table~\ref{tab:study2-clustering}. Under the uncontaminated setting, both methods achieved excellent first-layer clustering performance, although RDMM already exhibited noticeably better complete-pathway recovery. As the contamination rate increased, the performance of DGMM deteriorated rapidly, as reflected by substantially lower median ARI values, markedly higher misclassification rates, and increased between-replication variability. In contrast, RDMM remained considerably more robust, maintaining relatively high ARI values and low misclassification rates even under severe contamination. Although some performance degradation was observed for RDMM as the contamination rate increased, the decline was gradual and substantially smaller than that of DGMM. Furthermore, RDMM generally exhibited narrower interquartile ranges across all contamination levels, indicating greater stability over repeated Monte Carlo replications. The superiority of RDMM was particularly evident for complete-pathway clustering, where the performance gap between the two methods widened steadily as the contamination rate increased. Overall, these results provide strong empirical evidence that incorporating pathway-specific latent scale variables substantially enhances both the robustness and stability of deep mixture modelling in the presence of heavy-tailed contaminated observations.

\section{Case study: Chowdary data}
\label{sec:chowdary}

The Chowdary dataset comprises gene expression profiles from 104 human cancer tissue samples, including 64 lymph node-negative breast tumors and 40 Dukes' B colon tumors. It was originally generated by \citet{chowdary2006prognostic} to investigate whether tissue preservation using RNAlater affects gene expression measurements by comparing paired snap-frozen and RNAlater-preserved specimens. The dataset is publicly available from the NCBI Gene Expression Omnibus (GEO) under accession GSE3726. Following the preprocessing protocol of \citet{de2008clustering}, 182 informative genes were selected for subsequent analysis. The resulting gene expression data exhibit pronounced skewness and contain numerous extreme observations, posing substantial challenges for conventional statistical and machine learning methods. Consequently, the Chowdary dataset has become a widely used benchmark for evaluating the robustness and effectiveness of clustering and classification methods designed for high-dimensional gene expression data.

Both DGMM and RDMM were fitted using two latent layers. In this case study, clustering performance was evaluated using the first-layer component assignments. Since the first layer represents the primary partition of the observations, the number of first-layer components was fixed at $K_1=2$, equal to the known number of cancer classes. Candidate architectures were then formed by varying the number of second-layer components, $K_2\in\{2,4\}$, and the latent dimensions
$(r_1,r_2)$. For each candidate architecture, ten random initialisations were used to reduce sensitivity to local optima. The retained fit was the one attaining the best likelihood-based criterion among the ten runs. Clustering performance was evaluated using ARI and MR, after optimal matching between the estimated first-layer cluster labels and the known tissue classes.

For RDMM, a common degrees-of-freedom parameter was assumed across all pathways. The fitted value was approximately $\widehat{\nu}=2.050$, which is close to the lower admissible bound used in estimation and indicates pronounced heavy-tailed behaviour. This estimate supports the use of a Student-$t$ specification for the Chowdary data. The Gaussian deep mixture model was noticeably more difficult to fit stably on this data set. Several DGMM runs exhibited numerical instability or convergence problems, which is consistent with the sensitivity of Gaussian latent-variable models to heavy-tailed observations and extreme expression values. By contrast, the latent scale mechanism in RDMM downweights observations that are poorly explained by a pathway, thereby providing greater numerical stability and increased robustness.

\begin{table}[!ht]
\centering
\caption{Best clustering performance on the Chowdary cancer data. The best value in each performance column is shown in bold.}
\label{tab:chowdary_best_models}
\begin{tabular}{llrr}
\toprule
Method & Selected model & ARI & MR \\
\midrule
MCLUST & VVI                         & 0.0657 & 0.3462 \\
MFA    & $q=3$                       & 0.5858 & 0.1154 \\
MCFA   & $q=1$                       & 0.6800 & 0.0865 \\
DGMM   & $(K_2,r_1,r_2)=(2,9,5)$    & 0.1417 & 0.2981 \\
RDMM   & $(K_2,r_1,r_2)=(2,6,2)$    & \textbf{0.8867} & \textbf{0.0288} \\
\bottomrule
\end{tabular}
\vspace{2mm}
\parbox{0.92\linewidth}{\footnotesize
\textit{Note:} MR denotes the misclassification rate. The DGMM and RDMM results are based on first-layer memberships. The models in this table are selected according to their external clustering performance rather than according to BIC.}
\end{table}

Table~\ref{tab:chowdary_best_models} compares the best clustering performance achieved by the benchmark Gaussian methods reported by \cite{baek2011mixtures} with the best DGMM and RDMM configurations obtained in the present study. RDMM achieved the highest ARI, $0.8867$, and the lowest MR, $0.0288$. It therefore substantially outperformed all Gaussian competitors. In particular, relative to DGMM, RDMM increased the ARI by $0.7450$ and reduced the misclassification rate from $0.2981$ to $0.0288$. This corresponds to a relative reduction of approximately $90.3\%$ in the misclassification rate. The result supports the use of Student-$t$ latent distributions for gene-expression data containing heavy tails and extreme observations.

To isolate the effect of replacing the Gaussian specification by the robust Student-$t$ specification, Table~\ref{tab:chowdary_matched} compares DGMM and RDMM under identical values of $K_2$, $r_1$ and $r_2$. BIC is reported using the \texttt{mclust} sign convention, so that larger values are preferred within a given model family. Because DGMM and RDMM are based on different component densities and have different parameterisations, their absolute BIC values should
not be interpreted as a direct cross-family performance measure.

\begin{table}[htbp]
\centering
\caption{Comparison of DGMM and RDMM under identical model architectures on the Chowdary cancer data. BIC follows the \texttt{mclust} convention, for which larger values are preferred.}
\label{tab:chowdary_matched}
\begin{tabular}{ccccrrrrrr}
\toprule
&&&& \multicolumn{3}{c}{DGMM} & \multicolumn{3}{c}{RDMM} \\
\cmidrule(lr){5-7}\cmidrule(lr){8-10}
$K_2$ & $r_1$ & $r_2$ & Pathways
& BIC & ARI & MR
& BIC & ARI & MR \\
\midrule
2 & 4  & 2 & 4 & -13258.23 & 0.0657 & 0.3462 &  4727.29 & 0.0657 & 0.3462 \\
2 & 5  & 3 & 4 &  -1157.98 & 0.0657 & 0.3462 &  8936.84 & 0.0657 & 0.3462 \\
2 & 6  & 2 & 4 &  -4579.31 & 0.0187 & 0.3942 & 27504.95 & \textbf{0.8867} & \textbf{0.0288} \\
2 & 6  & 4 & 4 &    481.92 & 0.0792 & 0.3365 & 24874.69 & \textbf{0.8867} & \textbf{0.0288} \\
2 & 7  & 3 & 4 &   2182.23 & 0.0675 & 0.3462 & 26858.96 & 0.8505 & 0.0385 \\
2 & 7  & 5 & 4 &   2024.62 & 0.0934 & 0.3269 & 22832.48 & 0.8505 & 0.0385 \\
2 & 8  & 2 & 4 &   3056.64 & 0.0566 & 0.3558 & 31715.10 & 0.8151 & 0.0481 \\
2 & 8  & 4 & 4 &   3842.65 & 0.0566 & 0.3558 & 29548.65 & 0.7803 & 0.0577 \\
2 & 9  & 3 & 4 &    446.52 & 0.0370 & 0.3750 & 30476.63 & 0.7803 & 0.0577 \\
2 & 9  & 5 & 4 &  -2660.65 & \textbf{0.1417} & \textbf{0.2981} & 29481.08 & 0.8151 & 0.0481 \\
2 & 10 & 4 & 4 &   5469.22 & 0.0370 & 0.3750 & 33556.56 & 0.7131 & 0.0769 \\
2 & 11 & 5 & 4 &   6827.77 & 0.0283 & 0.3846 & 33850.45 & 0.7803 & 0.0577 \\
\midrule
4 & 4  & 2 & 8 & -13061.87 & 0.0657 & 0.3462 &  5468.55 & 0.0657 & 0.3462 \\
4 & 5  & 3 & 8 &  -1171.41 & 0.0657 & 0.3462 & 23708.50 & \textbf{0.8867} & \textbf{0.0288} \\
4 & 6  & 2 & 8 &  -4930.68 & 0.0187 & 0.3942 & 28110.07 & \textbf{0.8867} & \textbf{0.0288} \\
4 & 6  & 4 & 8 &    249.69 & 0.0548 & 0.3558 & 25400.89 & \textbf{0.8867} & \textbf{0.0288} \\
4 & 7  & 3 & 8 &  -4110.91 & 0.0116 & 0.4038 & 28852.63 & 0.8505 & 0.0385 \\
4 & 7  & 5 & 8 &   1413.42 & 0.0447 & 0.3654 & 23315.50 & 0.8505 & 0.0385 \\
4 & 8  & 2 & 8 &   4223.50 & 0.0566 & 0.3558 & 31719.71 & 0.8151 & 0.0481 \\
4 & 8  & 4 & 8 &   4221.07 & 0.0566 & 0.3558 & 27925.48 & 0.8151 & 0.0481 \\
4 & 9  & 3 & 8 &  -4985.05 & 0.0370 & 0.3750 & 32451.77 & 0.7803 & 0.0577 \\
4 & 9  & 5 & 8 &  -5363.53 & 0.0966 & 0.3269 & 31709.19 & 0.7803 & 0.0577 \\
4 & 10 & 4 & 8 &   6626.69 & 0.0370 & 0.3750 & 31562.77 & 0.7463 & 0.0673 \\
4 & 11 & 5 & 8 &   5633.22 & 0.0480 & 0.3654 & 34390.24 & 0.7803 & 0.0577 \\
\bottomrule
\end{tabular}%
\end{table}

The matched-architecture results reveal that RDMM generally provides a substantial improvement over DGMM. The only configurations for which the two methods produced essentially identical clustering were those with the smallest latent dimensions, such as $(r_1,r_2)=(4,2)$ and $(5,3)$. These architectures appear to impose such severe dimension reduction that the additional robustness of the Student-$t$ distribution cannot compensate for insufficient latent representation capacity. Once the first-layer latent dimension reached six or more, RDMM produced markedly higher ARI values and lower misclassification rates in almost every configuration.

The BIC-optimal architecture did not coincide with the architecture having the best external clustering performance. Under the \texttt{mclust} convention, the largest DGMM BIC was attained by
$(K_2,r_1,r_2)=(2,11,5)$, for which the ARI was only $0.0283$. The largest RDMM BIC was attained by $(4,11,5)$, with ARI $0.7803$, whereas several simpler architectures achieved the maximum RDMM ARI of $0.8867$. This discrepancy does not indicate a sign error in BIC. Rather, BIC balances maximised likelihood against model complexity, while ARI measures agreement with external class labels. The two criteria therefore optimise different objectives. A similar phenomenon was reported for the same Chowdary data by \cite{baek2011mixtures}, who found that BIC tended to select overly complex factor-analytic mixture models and did not identify the factor dimension associated with the best ARI and error rate. Accordingly, BIC remains useful for likelihood-based selection within a model family, but it should not be interpreted as a guarantee of optimal recovery of the biological classes.

To sum up, this case study provides two complementary findings. First, RDMM achieved substantially better first-layer clustering than DGMM and the Gaussian benchmark methods. Second, the robust Student-$t$ formulation improved the numerical behaviour of the deep mixture model in a data set exhibiting pronounced heavy tails and extreme observations. The consistently small estimate of the common degrees-of-freedom parameter further confirms that robustness is an essential feature for modelling these gene-expression data. An additional evaluation of the proposed RDMM on the Wine data set is provided in Appendix~\ref{appendix:wine}.

\section{Conclusions}
\label{sec:conclusions}

This paper has proposed a robust deep mixture model that integrates hierarchical deep mixture modelling with robust latent-variable modelling through a pathway-wise shared scale-mixture construction. Conditional on a selected pathway, a single observation-level latent precision variable is shared across the deepest latent distribution, every intermediate latent transition, and the observation model, yielding an exact multivariate Student-$t$ distribution for each complete pathway. This shared-scale formulation enables atypical observations to be coherently down-weighted throughout the entire latent hierarchy while preserving the hierarchical representation, dimension-reduction capability, and parsimonious parameter-sharing structure of the DGMM. A stochastic expectation--maximisation algorithm was developed for parameter estimation, and the numerical studies demonstrate that the proposed framework accurately estimates the degrees-of-freedom parameters and consistently improves clustering performance over the DGMM under heavy-tailed and contaminated settings.

The proposed framework nevertheless has several limitations. As the depth of the hierarchy and the number of mixture components increase, the number of complete pathways grows rapidly, leading to increased computational cost. In addition, pathway-specific degrees-of-freedom parameters may become unstable when effective pathway sample sizes are small, and the stochastic estimation procedure remains sensitive to initialization and Monte Carlo variability. Future work will investigate pathway pruning, regularised or shared degrees-of-freedom parameterisations, and more scalable variational or Bayesian inference. Extensions to mixed-type data, skewed latent distributions, and more general robust deep probabilistic mixture models also represent promising directions for future research.

\appendix

\section{Additional real data results}
\label{appendix:wine}

The Wine data set consists of measurements on 27 chemical and physical properties of wines produced in the Piedmont region of Italy. The data comprise three well-separated cultivars: Barolo (59 observations), Grignolino (71 observations), and Barbera (48 observations). Owing to its clear cluster structure, the Wine data set has become a standard benchmark for evaluating clustering methods. To further evaluate the proposed RDMM, we considered the Wine data set and compared its clustering performance with several existing methods. For consistency with the literature, the performance results for PAM, hierarchical clustering, Gaussian mixture models (GMM), skew-normal mixture models (SNmm), skew-$t$ mixture models (STmm), factor mixture analysis (FMA), mixture of factor analysers (MFA), and DGMM were taken directly from \citet{viroli2019deep}. For RDMM, a range of model configurations was examined by varying the numbers of first- and second-layer mixture components ($k_1$ and $k_2$) together with the corresponding latent dimensions ($r_1$ and $r_2$). The optimal RDMM configuration was selected as $k_1=3$, $k_2=4$, $r_1=4$, and $r_2=1$. Since the Wine data contain three known cultivars, observations were assigned to first-layer clusters according to the maximum marginal first-layer posterior probabilities. Clustering performance was assessed using ARI and the MR. The results are reported in Table~\ref{tab:wine_compare}.

\begin{table}[!ht]
\centering
\caption{Comparison of clustering performance on the Wine data. Results for the competing methods are taken from \citet{viroli2019deep}; the RDMM result is obtained in the present study.}
\label{tab:wine_compare}
\begin{tabular}{lcc}
\hline
Method & ARI & MR \\
\hline
PAM & 0.863 & 0.045 \\
Hierarchical clustering & 0.865 & 0.045 \\
GMM & 0.917 & 0.028 \\
SNmm & 0.964 & 0.011 \\
STmm & 0.085 & 0.511 \\
FMA & 0.361 & 0.303 \\
MFA & 0.983 & 0.006 \\
DGMM & 0.983 & 0.006 \\
\textbf{RDMM} & \textbf{1.000} & \textbf{0.000} \\
\hline
\end{tabular}
\end{table}

As shown in Table~\ref{tab:wine_compare}, the proposed RDMM achieves perfect clustering performance, with an ARI of 1.000 and an MR of 0.000, outperforming all competing methods. Although MFA and DGMM already achieve excellent performance on this benchmark dataset, both methods still misclassify one observation (MR = 0.006). In contrast, RDMM correctly classifies all 178 wine samples, demonstrating that the proposed pathway-wise shared scale-mixture formulation can further improve clustering accuracy on well-separated data.

\bibliographystyle{unsrtnat}
\bibliography{sn-bibliography}

@article{fokoue2003mixtures,
  title={Mixtures of factor analysers. {B}ayesian estimation and inference by stochastic simulation},
  author={Fokou{\'e}, Ernest and Titterington, DM},
  journal={Machine Learning},
  volume={50},
  number={1},
  pages={73--94},
  year={2003},
  publisher={Springer}
}

@article{mclachlan2003modelling,
  title={Modelling high-dimensional data by mixtures of factor analyzers},
  author={McLachlan, Geoffrey J and Peel, David and Bean, Richard W},
  journal={Computational Statistics \& Data Analysis},
  volume={41},
  number={3-4},
  pages={379--388},
  year={2003},
  publisher={Elsevier}
}

@article{mclachlan2007extension,
  title={Extension of the mixture of factor analyzers model to incorporate the multivariate $t$-distribution},
  author={McLachlan, Geoffrey J and Bean, Richard W and Jones, L Ben-Tovim},
  journal={Computational Statistics \& Data Analysis},
  volume={51},
  number={11},
  pages={5327--5338},
  year={2007},
  publisher={Elsevier}
}

@article{viroli2019deep,
  title={Deep {G}aussian mixture models},
  author={Viroli, Cinzia and McLachlan, Geoffrey J},
  journal={Statistics and Computing},
  volume={29},
  number={1},
  pages={43--51},
  year={2019},
  publisher={Springer}
}

@article{wang2026robust,
  title={Robust mixture modeling using the truncated $t$ distribution: {R}ecent advances and new results},
  author={Wang, Wan-Lun and Lee, Si-Hui and Lin, Tsung-I},
  journal={Advances in Data Analysis and Classification},
  pages={1--28},
  year={2026},
  publisher={Springer}
}

@article{kock2022variational,
  title={Variational inference and sparsity in high-dimensional deep {G}aussian mixture models},
  author={Kock, Lucas and Klein, Nadja and Nott, David J},
  journal={Statistics and Computing},
  volume={32},
  number={5},
  pages={70},
  year={2022},
  publisher={Springer}
}

@inproceedings{fuchs2022mi2ami,
  title={{MI2AMI: M}issing data imputation using mixed deep {G}aussian mixture models},
  author={Fuchs, Robin and Pommeret, Denys and Stocksieker, Samuel},
  booktitle={{International Conference on Machine Learning, Optimization, and Data Science}},
  pages={211--222},
  year={2022},
  organization={Springer}
}

@article{gorshenin2025small,
  title={Small sample classification and regression for time series and tabular data based on deep {G}aussian mixture models},
  author={Gorshenin, AK and Dostovalova, AM},
  journal={Pattern Recognition and Image Analysis},
  volume={35},
  number={2},
  pages={83--93},
  year={2025},
  publisher={Springer}
}

@article{viroli2021deep,
  title={Deep mixtures of unigrams for uncovering topics in textual data},
  author={Viroli, Cinzia and Anderlucci, Laura},
  journal={Statistics and Computing},
  volume={31},
  number={3},
  pages={22},
  year={2021},
  publisher={Springer}
}

@inproceedings{selosse2020bumpy,
  title={A bumpy journey: {E}xploring deep {G}aussian mixture models},
  author={Selosse, Margot and Gormley, Claire and Jacques, Julien and Biernacki, Christophe},
  booktitle={{``I Can't Believe It's Not Better!'' NeurIPS 2020 Workshop}},
  year={2020}
}

@article{hamalainen2020deep,
  title={Deep residual mixture models},
  author={H{\"a}m{\"a}l{\"a}inen, Perttu and Trapp, Martin and Saloheimo, Tuure and Solin, Arno},
  journal={arXiv preprint arXiv:2006.12063},
  year={2020}
}

@article{fuchs2022mixed,
  title={Mixed deep {G}aussian mixture model: {A} clustering model for mixed datasets},
  author={Fuchs, Robin and Pommeret, Denys and Viroli, Cinzia},
  journal={Advances in Data Analysis and Classification},
  volume={16},
  number={1},
  pages={31--53},
  year={2022},
  publisher={Springer}
}

@article{baek2011mixtures,
  title={Mixtures of common $t$-factor analyzers for clustering high-dimensional microarray data},
  author={Baek, Jangsun and McLachlan, Geoffrey J},
  journal={Bioinformatics},
  volume={27},
  number={9},
  pages={1269--1276},
  year={2011},
  publisher={Oxford University Press}
}

@article{peel2000robust,
  title={Robust mixture modelling using the $t$ distribution},
  author={Peel, David and McLachlan, Geoffrey J},
  journal={Statistics and Computing},
  volume={10},
  number={4},
  pages={339--348},
  year={2000},
  publisher={Springer}
}

@article{mclachlan2019finite,
  title={Finite mixture models},
  author={McLachlan, Geoffrey J and Lee, Sharon X and Rathnayake, Suren I},
  journal={Annual Review of Statistics and its Application},
  volume={6},
  number={1},
  pages={355--378},
  year={2019},
  publisher={Annual Reviews}
}

@article{wang2022robust,
  title={Robust clustering of multiply censored data via mixtures of $t$ factor analyzers},
  author={Wang, Wan-Lun and Lin, Tsung-I},
  journal={Test},
  volume={31},
  number={1},
  pages={22--53},
  year={2022},
  publisher={Springer Nature BV}
}

@inproceedings{rakotonirina2026unsupervised,
  title={Unsupervised spatially aware {G}aussian mixture model via implicit deep priors},
  author={Rakotonirina, Herbert and Lohier, Th{\'e}ophile and Baptiste, Julien},
  booktitle={{Proceedings of the IEEE/CVF Winter Conference on Applications of Computer Vision}},
  pages={749--758},
  year={2026}
}

@article{mahdavi2026image,
  title={Image segmentation using logit-$t$ mixtures},
  author={Mahdavi, Abbas and Contreras-Reyes, Javier E},
  journal={Statistical Methods \& Applications},
  pages={1--27},
  year={2026},
  publisher={Springer}
}

@book{goodfellow2016deep,
  author    = {Goodfellow, Ian and Bengio, Yoshua and Courville, Aaron},
  title     = {{Deep Learning}},
  year      = {2016},
  publisher = {MIT Press},
  address   = {Cambridge, MA}
}

@article{lecun2015deep,
  title={Deep learning},
  author={LeCun, Yann and Bengio, Yoshua and Hinton, Geoffrey},
  journal={Nature},
  volume={521},
  number={7553},
  pages={436--444},
  year={2015},
  publisher={Nature Publishing Group UK London}
}

@book{mclachlan2000finite,
  title={{Finite Mixture Models}},
  author={McLachlan, Geoffrey J and Peel, David},
  year={2000},
  address   = {Hoboken, NJ},
  publisher={John Wiley \& Sons}
}

@inproceedings{mclachlan1998robust,
  title={Robust cluster analysis via mixtures of multivariate $t$-distributions},
  author={McLachlan, Geoffrey J and Peel, David},
  booktitle={{Joint IAPR International Workshops on Statistical Techniques in Pattern Recognition (SPR) and Structural and Syntactic Pattern Recognition (SSPR)}},
  pages={658--666},
  year={1998},
  organization={Springer}
}

@article{nielsen2000stochastic,
  title={The Stochastic {EM} Algorithm: {E}stimation and Asymptotic Results},
  author={Nielsen, S{\o}ren Feodor},
  journal={Bernoulli},
  pages={457--489},
  year={2000},
  publisher={JSTOR}
}

@article{levine2001implementations,
  title={Implementations of the {M}onte {C}arlo {EM} algorithm},
  author={Levine, Richard A and Casella, George},
  journal={Journal of Computational and Graphical Statistics},
  volume={10},
  number={3},
  pages={422--439},
  year={2001},
  publisher={Taylor \& Francis}
}

@article{sheng2026robust,
  title={Robust Adaptive Filtering via Maximum Likelihood Method Based on Student's-$t$ Mixture Model},
  author={Sheng, Lingjie and Chien, Ying-Ren and Qian, Junhui and Qian, Guobing and Wang, Shiyuan},
  journal={IEEE Signal Processing Letters},
  volume={33},
  pages={1096--1100},
  year={2026},
  publisher={IEEE}
}

@article{de2008clustering,
  title={Clustering cancer gene expression data: {A} comparative study},
  author={De Souto, Marcilio CP and Costa, Ivan G and De Araujo, Daniel Sa and Ludermir, Teresa B and Schliep, Alexander},
  journal={BMC Bioinformatics},
  volume={9},
  number={1},
  pages={497},
  year={2008},
  publisher={Springer}
}

@article{chowdary2006prognostic,
  title={Prognostic gene expression signatures can be measured in tissues collected in {RNA}later preservative},
  author={Chowdary, Dondapati and Lathrop, Jessica and Skelton, Joanne and Curtin, Kathleen and Briggs, Thomas and Zhang, Yi and Yu, Jack and Wang, Yixin and Mazumder, Abhijit},
  journal={The Journal of Molecular Diagnostics},
  volume={8},
  number={1},
  pages={31--39},
  year={2006},
  publisher={Elsevier}
}

\end{document}